# Abstracting data in distributed ledger systems for higher level analytics and visualizations

**Leny Vinceslas**
Loughborough University London

**Hirsh Pithadia**
University College London & RegulAItion Ltd.

**Safak Dogan**
Loughborough University London

**Srikumar Sundareshwar**
RegulAItion Ltd.

**Ahmet M. Kondoz**
Loughborough University London

*Abstract*—By design, distributed ledger technologies persist low-level data which makes conducting complex business analysis of the recorded operations challenging. Existing blockchain visualization and analytics tools such as block explorers tend to rely on this low-level data and complex interfacing to provide enriched level of analytics. The ability to derive richer analytics could be improved through the availability of a higher level abstraction of the data. This article proposes an abstraction layer architecture that enables the design of high-level analytics of distributed ledger systems and the decentralized applications that run on top. Based on the analysis of existing initiatives and identification of the relevant user requirements, this work aims to establish key insights and specifications to improve the auditability and intuitiveness of distributed ledger systems by leveraging the development of future user interfaces. To illustrate the benefits offered by the proposed abstraction layer architecture, a regulated sector use case is explored.

■ **DISTRIBUTED LEDGER TECHNOLOGIES (DLTs)** are becoming more widely used. They record interactions between multiple parties in an immutable way. They are built on consensus based decentralized smart contract systems which address the trust issue between the involved parties. Numerous applications of distributed ledgers are currently being developed in various fields of the industry such as agriculture, finance, security, digital property or healthcare.[1] Hyperledger Fabric (HF) is a permissioned blockchain framework targeting enterprise-grade business applications in which only authorized organizations can access the ledger.[2] By design, they persist low-level data that makes conducting complex business analysis of the recorded operations challenging.

For instance, in HF the record of operations can be accessed usually by third party applications via querying the ledger. This is achieved by employing a native set of Application Programming Interfaces (APIs) where information about transactions, smart contracts or blocks can only be queried by their corresponding cryptographic hash. Although this access method allows for data lookup, such basic APIs are not adequate to devise high-level information from blockchains, often needed for analytics. Consequently, block explorers and similar visualization and analytics tools often only offer limited unintuitive information. These challenges call for an innovative approach for introducing a middleware between the presentation and data query layers that enables more accessible analytics and information visualizations. This can be achieved by employing an abstraction layer which aggregates data from the ledger, pre-processes it and provides higher level APIs to block explorers and analytics dashboards that can then intuitively present information readily.

This article proposes an abstraction layer framework that facilitates better design of business analytics for DLT based systems and the Decentralized Applications (DApps) that run on them. The purpose of this work is to establish a higher level of abstraction that improves the auditability and intuitiveness of distributed ledger records and enables further development of future user interfaces including analytics and visualization tools. Based on the analysis of existing initiatives and identification of the relevant user requirements, we infer specifications to improve the auditability and usability of block explorers. An abstraction layer has been designed to bridge the gap between a blockchain ledger and a user interface. As a result, the proposed abstraction layer coupled with a new User Interface (UI) promotes the ease of analytics such as tracking and tracing of the history of operations, clustering of user addresses, and labeling of entities. Finally, to illustrate the benefits offered by the proposed abstraction layer architecture, an industrial case study anchored in regulated sectors is explored.

The article is organized as follows. We first analyze the challenges and limitations related to designing intuitive visualizations for DLTs. Secondly, we review the related work by highlighting notable aspects in existing block explorers and abstraction layer implementations. Thirdly, we propose an approach to build account and transaction-oriented abstractions. We then provide an example of application in a case study. Finally, a summary of the article is provided with an outlook on future of the topic.

CHALLENGES AND LIMITATIONS

In data analytics, visual representations are primarily designed to make sense of data and provide insights. They usually model data structures, which help to expand the boundary of individuals' cognitive system.[3] Visual representations not only support users' reasoning, but also provide a construct to manipulate information. They can be used to both structure information and reduce individuals' cognitive burden by easing external anchoring, information foraging, and cognitive offloading.[4]

In DLTs, audits are often conducted through lists and tables of low-level block data that do not easily allow for tracking and tracing digital assets.[5] The introduction of adequate visual representations of the ledgers' data has the capacity to enable a higher level of functionalities, and therefore improve the intuitiveness of auditing processes.

To build intuitive visualizations or conduct in-depth analysis of blockchain's recorded operations, data must be readily available in a format that can be consumed by the frontend visual representations. Although most DLTs provide an access to the ledger via their Software Development Kits (SDKs), they only offer low-level query interfaces that often lack in semantic richness or functionality. For instance, information about transactions, contracts or blocks can usually be queried by their corresponding hashes. Such basic query interfaces are not adequate to map out high-level information or derive visualizations from blockchains' ledgers, and hence make it significantly challenging for users to deduct valuable insights that can serve complex business analysis quickly.

As a result, a few block explorers in literature and marketplace implement high-level ledger visualizations or analysis functionalities. These limitations drastically reduce the possibility to adapt the abstraction of the displayed information to the targeted audience and application. The level of abstraction needs careful consideration when designing a visualization system to avoid situations where the users are presented with either insufficient or too much low-level data. In visualization and analytics, mismatched level of granularity can result in both less accurate comprehension of a situation and higher time to complete a given task.[6]

Curating information and presenting users with notable insights at the right level of granularity is often the responsibility of data analysts and designers. However, in the relatively new DLTs landscape, there are no such easily accessible solutions to implement the appropriate blockchain visualization tools to address a target audience's needs in a specific scenario.[6]

The lack of high-level abstraction makes the development of analytics time consuming for the developers since new functionalities need to be designed and implemented to analyze and aggregate the available low-level data. Additionally, delegating such functionalities to frontend applications can result in high resource-consuming services for the devices rendering such information. Therefore, there is a clear need for a standardized intermediate-level abstraction that provides higher level information from the low-level block data.[7]

BACKGROUND AND RELATED WORK

In DLTs, the ledger is a global data structure collectively maintained by a set of mutually untrusting participants.[8] Changes to the ledger are organized into transactions which record the identifiers of their creators and beneficiaries. Transactions are hashed and grouped into blocks which are then chained together. Each block is appended via its header pointing to its predecessor. The synchronization of all peers on the state of the blockchain network is achieved using a consensus algorithm. This append-only ledger system provides to DLTs immutable records of transactions and therefore makes the blockchain tamper-resistant. In addition to transactions, DLTs can implement smart contracts. Smart contracts are executable scripts that read or write to the ledger and are deployed across peers of the network.

A variety of block explorers have been developed to visualize and analyze the transaction details and network activity of different distributed ledger platforms. These often fail to provide DApp-specific information easily.

For example, on its main view Hyperledger Explorer[9] presents users with statistical insights on the HF network (number of blocks, transactions, nodes and smart contracts) grouped by organizations or averaged over time. It also displays the name of the peers operating on that specific channel and details about the last committed blocks.

Alethio block explorer[10] provides richer analytics. In addition to the standard browsing history features, it maps interactions between accounts by tracing transactions and evoked smart contracts. These interactions are visually represented using simple node-link diagrams. It also allows to keep track of account balances, search information by account alias rather than by block and transaction hashes, and attach diverse social information to account addresses. Its functionalities, such as address tracking, tracing, labeling and data aggregation of DLT data improve auditability.

Motivated by the need of providing secure and decentralized services, DLTs keep track of very large amounts of ever-growing data.[5] To overcome these limitations, both abovementioned block explorers implement an additional backend software sitting between the UI and ledger's low-level query interface. This standardized middleware aims to abstract the complexity of user interactions with blockchains, and is responsible for querying, aggregating, and conditioning the ledger data, so that it is capable of offering higher level analytics more easily.

Several platforms have advocated for the need for abstraction layers. For instance, Ledgerdata Refiner[7] is a ledger data query platform developed for interfacing permissioned DLTs such as HF. It is based on a data analysis middleware which extracts and synchronizes the ledger data, and then parses the relationship among them. From the queried blocks and transactions, the middleware provides end users with tailored queries to access aggregated ledger information.

Datachain is another example, which is an interoperable framework that eases the extraction of data from different underlying blockchains.[11] It allows users to define specific high-level query abstractions, and perform data requests, extract transactions, manage data assets and derive high-level analytic insights automatically.

PROPOSED APPROACH

We propose two types of visual representations: (i) a transaction-oriented abstraction emphasizing on the time series of the transaction history while allowing tracing and tracking of assets, and (ii) an account-oriented abstraction focusing on interactions between entities of the audited DLTs and providing insights on inter-party behaviors.

The transaction-oriented abstraction uses a directed acyclic graph layout. As shown in Figure 1a, vertices represent transactions while directed edges illustrate the transaction flow between the source outputs and target inputs. This visualization shows the flow of transactions relative to a given transaction. It allows tracking and tracing of assets from their origins to end points across a predefined number of hops. For a low granularity level, only one-hop tracking is displayed with respect to $Tx_3$ which corresponds to the blue vertices in Figure 1a. For a higher granularity level a two-hop tracking is represented with blue and grey vertices. Using an adaptive design that displays details on demand, this visualization gives access to a continuum of granularity. In addition to the vertices,

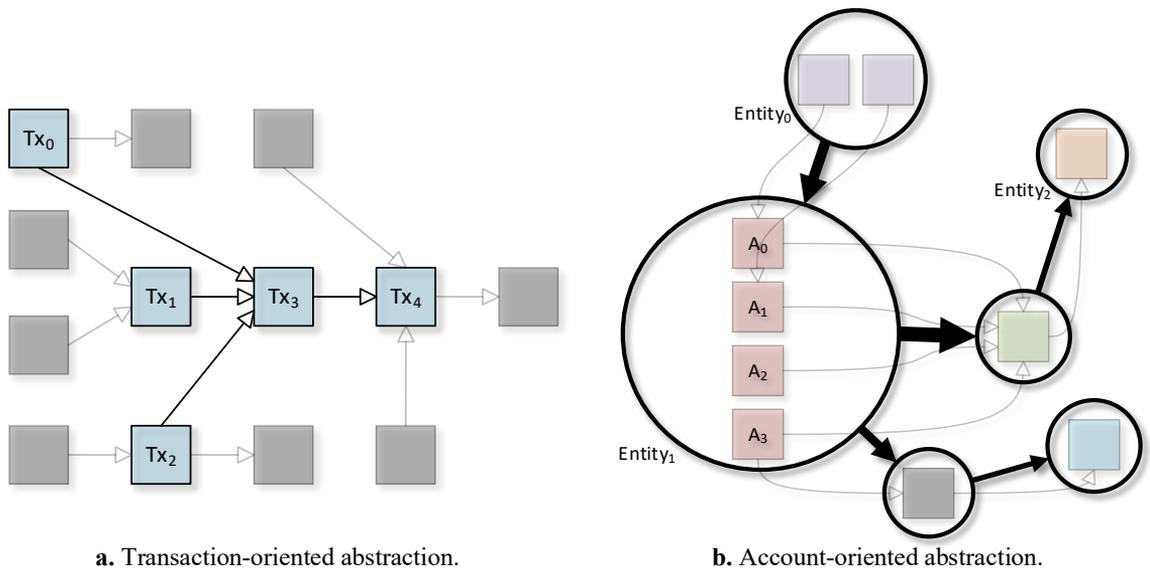

**a.** Transaction-oriented abstraction.   **b.** Account-oriented abstraction.

**Figure 1.** Abstracted visualizations.

the directed edges can be augmented with asset values and transaction timestamps. Informing about the smart contracts that triggered the represented transactions can also provide pertinent insights. The implementation of this transaction-oriented abstraction requires knowledge about the mapping between the transactions of interest. However, this information is not directly available in the ledger and must be obtained through data parsing, aggregation and analysis.

The account-oriented abstraction uses a force-directed graphs where different granularity levels are implemented. As shown in Figure 1b, when a macro level is chosen, the square-shaped vertices represent addresses while directed edges are illustrating interactions between accounts. At a lower level, circular vertices denote clusters of accounts, forming entities linked by the directed edges. Entities and directed edges can be of variable sizes, representing the quantity of accounts by cluster and total value or amount of all inter-cluster interactions that occurred during a predefined time period respectively. By nesting accounts within different cluster sizes, the visualization can efficiently adapt to the required level of detail. The implementation of the account-oriented abstraction requires knowledge on entities and their interactions, which needs clustering and labeling the ledger data.

To address the implementation of the two visual representations, we propose a general architecture for building an abstraction layer. The main purpose of this layer is to hide the complex interactions with ledgers.

Through a simplified interface, users can connect to underlying ledgers to derive high-level analytic insights or perform high-level requests such as asset tracking and tracing. With this abstraction layer, we establish a framework where different services can be easily integrated to provide a transparent and richer query interface for business analytics. As depicted in Figure 2, it first extracts transaction, block, contract and channel details through blockchain ledger SDKs. At this stage, it is conceivable to query data from different DLTs. The ledger data is then parsed and aggregated to create comprehensive objects with common data structures, easy to manipulate in a given framework. The parsed data is then cached so that both current and historic data can be accessed by the pre-processing services. Pre-processing services aim at deriving high-level information from the low-level cached data. For instance, this is where information is mapped or filtered according to predefined heuristics. The last service of the abstraction layer provides interfaces to query the computed analytics.

**Parsing** is the process of converting raw low-level data structures into higher level objects. Blockchain data structures are optimized for transaction validations and data retrieval across a distributed network and thus are not best suited for conducting analysis easily. For instance, to implement the proposed transaction-oriented abstraction, the parsing procedure must first collect the transactions from one or several blocks prior to mapping their inputs to previous transaction outputs. In addition, transactions must be assigned with IDs and timestamps along with the associated addresses of creators and beneficiaries to facilitate retrieval procedures.[12] Due to their inherent differences, Unspent Transaction Outputs (UTXOs) model and account model-based ledgers require notably different parsing procedure.

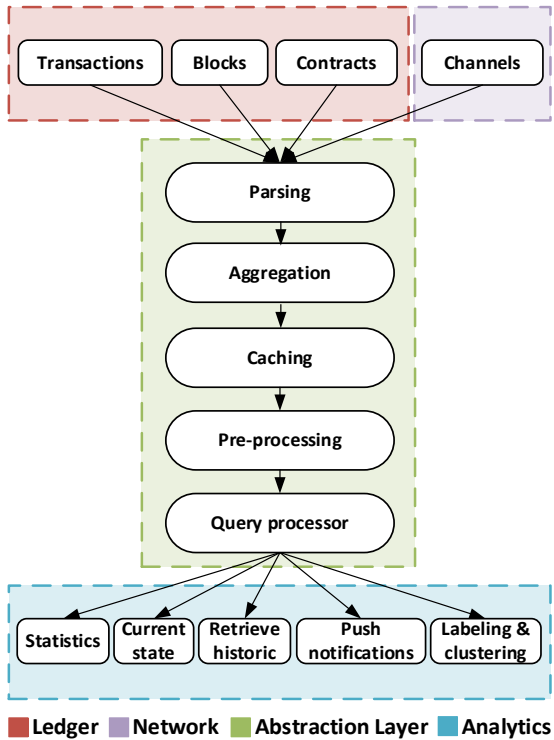

**Figure 2.** Abstraction layer architecture.

**Aggregation** refers to the collection and integration of data from multiple sources into a single storage destination. During this process, the different data sources required to infer higher level information are gathered and stored within a common data structure. For example, the proposed transaction-oriented abstraction needs to establish links between transaction inputs and outputs. To derive this mapping, transaction metadata of different blocks is aggregated, and corresponding source and destination addresses are matched.

**Caching** is the process of storing data resulting from previous computations so that future requests for that data can be executed faster. Both hardware and software used for caching depend on critical requirements such as the data volume, persistence time, access rate, throughput, and format. In the present scenario, the parsed and aggregated data can be cached in a server hard drive and RAM using a regular or graph database. The latter usually provides better basis for analyzing relationships between entities.[13]

**Pre-processing** defines the operation of taking the cached data as input to generate the information requested by the query services. For example, this step is necessary to compute statistical insights on the network state, i.e., number of transactions per day. In addition to cached data, the pre-processing operation can also request data from third-party services. In the case of the proposed account-oriented abstraction, a pre-processing service will access the stored aggregated data to cluster addresses based on various possible heuristics.[14] Entities can then be inferred from the clustered accounts. Address clustering is particularly powerful when combined with labeling, i.e., labeling clusters with real-world entity designations.[15] At a small scale, labels can be determined by users through the query services. However, for large-scale labeling, automated scraping of open-source information or access to a third-party service provider is desirable.

**Query processor** refers to the interfaces allowing third-party applications or users to query high-level data through a set of predefined instructions. Queries can initiate reading pieces of information collected or generated by the other abstraction layer services. Through a set of rich queries, this service aims to deliver requested data in a readily consumable format. To build the proposed visualizations, the query services can be implemented using Representational State Transfer (REST) APIs and the JSON file format. The defined set of APIs will allow client applications to remotely execute pre-processing services to submit labels and clustering rules before querying the pre-processed data.

USE CASE SCENARIO

RegNet is a privacy-preserving data-access and data-collaboration platform for the regulated sectors and addresses data-privacy challenges by combining DLTs, cryptography and machine learning. Participants can request and provide access to each other's data while resulting sharing agreements are stored in a HF platform. RegNet seeks to provide a trusted infrastructure to enable the exchange of data in a way where no sensitive information leaves the data-holders' firewalls. To ensure both security and relevance of exchanged information, RegNet uses privacy enhancing techniques on the data accessed between participants. In addition, RegNet also implements federated learning capabilities that promotes a secured collaborative way to build larger data models together with semi-trusted participants.

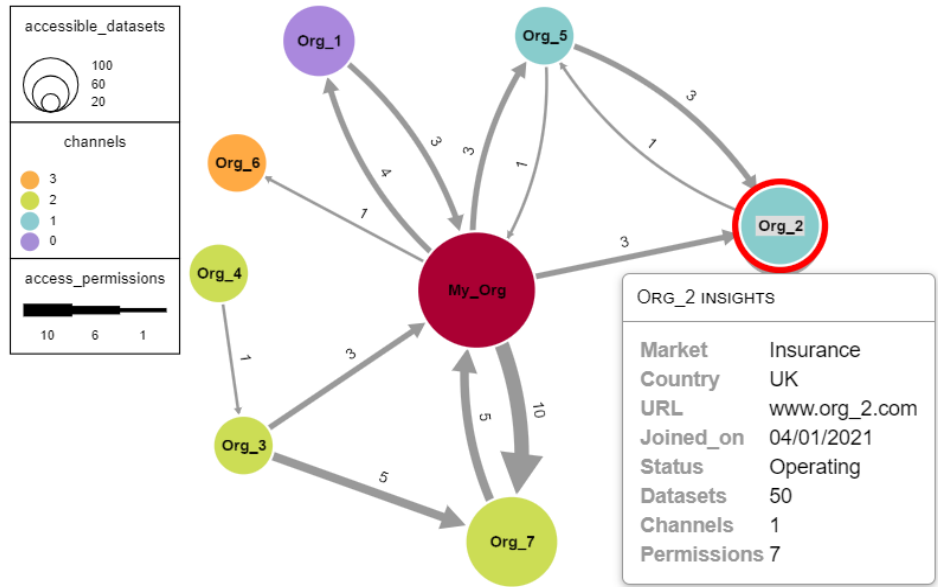

**Figure 3.** RegNet network visualization.

When building a suitable auditing layer for the data access applications that run on RegNet, a clear need for abstraction from block data was recognized. We therefore designed and implemented the discussed abstraction layer to more accessibly provide the interfacing needed for analytics.

Figure 3 shows the implementation of one such account-oriented dataset access based analytics utility onto the RegNet platform. It features nodes that represent organizations and edges that specify the relationships between nodes. Nodes are clustered accounts belonging to a same organization and thus sum up all account activities of a participant. The number of data models made available by each organization is reflected by the node sizes while the node color is indicating the channel on which an organization operates. In HF, channels are separated ledgers which enable the privacy and the scalability of the platform. The width of the directed edges illustrates the quantity of access permission granted between two organizations which is also numerically displayed. Adaptive granularity is introduced by a tooltip providing a summary and further insights onto an organization when its node is double-clicked.

## CONCLUSION AND OUTLOOK

In data analytics, visual representations are important since they can provide constructs that intuitively assist with inferring new information and can also reduce individuals' cognitive burden. The introduction of adequate visual representations for ledger data enables higher level of analytics and therefore augments the intuitiveness of auditing processes. Nevertheless, when designing analytics solutions, the level of abstraction needs careful consideration; higher level data can facilitate richer and quicker analytics.

To apply these concepts, we designed two distributed ledger visual representations, i.e., a transaction-based and an account-based abstractions. To enable the design of these high-level visual representations, we proposed an abstraction layer architecture. To illustrate the proposed visual concepts and application architecture, a use case based on an application for the regulated sector has been explored. Ultimately, a universal higher level query language could be designed on top of this abstraction layer which sits on blockchains. In turn, just like what Structured Query Language (SQL) is to Relational Database Management System (RDBMS), such language with its compositional, pragmatic and rich semantics would make business level querying much easier.

ACKNOWLEDGMENT

This work was supported by Innovate UK [application number 45079; project number 106159].